\documentclass[12pt,preprint]{aastex}

\usepackage[]{natbib}
\usepackage{fancyhdr,lastpage,textcomp,wrapfig,lipsum}
\usepackage[]{natbibspacing}
\usepackage{color}
\usepackage[dvipsnames]{xcolor}
\usepackage{enumitem}
\usepackage[colorlinks,urlcolor=roman_avg,citecolor=roman_avg,linkcolor=roman_avg]{hyperref}
\usepackage{amsmath}  
\usepackage{lmodern}
\usepackage{sansmath}
\usepackage[T1]{fontenc}
\usepackage{array}
\usepackage{floatrow}
\usepackage{wrapfig}
\usepackage{sidecap}
\usepackage{tocloft}  
\usepackage{tcolorbox}
\usepackage{apjfonts}
\usepackage{pdfpages}
\usepackage{float}
\definecolor{roman}{HTML}{70cbff} 
\definecolor{roman_dark}{HTML}{00276b} 
\definecolor{roman_avg}{HTML}{3879B5}

\usepackage{etoolbox}
\makeatletter
\patchcmd{\hyper@makecurrent}{%
    \ifx\Hy@param\Hy@chapterstring
        \let\Hy@param\Hy@chapapp
    \fi
}{%
    \iftoggle{inappendix}{
        \@checkappendixparam{chapter}%
        \@checkappendixparam{section}%
        \@checkappendixparam{subsection}%
        \@checkappendixparam{subsubsection}%
        \@checkappendixparam{paragraph}%
        \@checkappendixparam{subparagraph}%
    }{}%
}{}{\errmessage{failed to patch}}

\newcommand*{\@checkappendixparam}[1]{%
    \def\@checkappendixparamtmp{#1}%
    \ifx\Hy@param\@checkappendixparamtmp
        \let\Hy@param\Hy@appendixstring
    \fi
}
\makeatletter

\newtoggle{inappendix}
\togglefalse{inappendix}

\apptocmd{\appendix}{\toggletrue{inappendix}}{}{\errmessage{failed to patch}}

\renewcommand{\baselinestretch}{1.1}
\title{Roman Observations Time Allocation Committee \\ Final Report and Recommendations}

\author{Gail Zasowski \& Saurabh W.~Jha (co-Chairs), \\ 
Laura~Chomiuk,
Xiaohui~Fan,
Ryan~Hickox,
Dan~Huber,
Eamonn~Kerins,
Chip~Kobulnicky,
Tod~Lauer,
Masao~Sako,
Alice~Shapley,
Denise~Stephens,
David~Weinberg, and
Ben~Williams \\ 
\vskip 0.1in
{\it Ex officio: Lee~Armus, Thomas~Barclay, Ori~Fox, Karoline~Gilbert, \\ Jeff~Kruk, Patrick~Lowrance, Julie~McEnery, and Kristen~McQuinn}\\
\vskip 0.1in
April 24, 2025}  

\begin{document}

\maketitle

\vspace{0.5in}
\renewcommand{\baselinestretch}{1.0}\normalsize
\tableofcontents
\renewcommand{\baselinestretch}{1.1}\normalsize

\clearpage

\vspace*{0.1in}
\section{Executive Summary}

\begin{tcolorbox}[width=\linewidth, colback=white!87!roman_avg, rightrule=2mm, leftrule=2mm, left=2mm, right=2mm, top=2mm, bottom=2mm]

The \emph{Nancy Grace Roman Space Telescope} is poised to revolutionize our scientific understanding of exoplanets, dark matter, dark energy, and general astrophysics, including through an innovative community approach to defining and executing sky surveys. The Roman Observations Time Allocation Committee (ROTAC) was convened to recommend time allocations for the three Core Community Surveys (CCS) using the Wide Field Instrument (WFI): the High Latitude Wide Area Survey (HLWAS), the High Latitude Time Domain Survey (HLTDS), and the Galactic Bulge Time Domain Survey (GBTDS), as well as balance the time allocation for the General Astrophysics Surveys (GAS). 

\vspace{12pt} Each CCS had a corresponding Definition Committee that collected community input and designed proposals for a nominal (in-guide) survey, as well as underguide and overguide options with smaller and larger time allocations, respectively. These options explored different ways of fulfilling the mission science requirements while maximizing general astrophysics science goals enabled by the surveys. 

\vspace{12pt} After a thorough review of these reports, the ROTAC recommends:
\begin{enumerate}[nolistsep]
    \item a nominal allocation (520~days) for the HLWAS (\autoref{sec:hlwas}, \autoref{app:hlwas}), 
    \item a nominal allocation (180~days) for the HLTDS (\autoref{sec:hltds}, \autoref{app:hltds}),
    \item an overguide allocation (438~days) for the GBTDS with modifications (\autoref{sec:gbtds}, \autoref{app:gbtds}), and
    \item 389~days (25.5\% of the science operations time) reserved for GAS, including the early-definition Galactic Plane Survey (\autoref{sec:gas}). 
\end{enumerate}
\end{tcolorbox}

\clearpage

\section{Overview of Roman and the Wide Field Instrument Observing Program}

The \emph{Nancy Grace Roman Space Telescope} was the top-ranked scientific priority of the 2010 Astronomy and Astrophysics Decadal Survey \citep{astro2010}; it targets the most compelling questions in contemporary astrophysics, including the demographics of extrasolar planetary systems and the nature of dark matter and dark energy, through wide-field, high-resolution near-infrared imaging from space. Roman's Wide Field Instrument (WFI) will have an expansive 0.28~deg$^2$ field of view, sensitivity and resolution comparable to Hubble, and observing modes that support survey operations roughly 1000~times more efficiently than Hubble. Addressing Roman's key science goals by leveraging its observational survey speed led to a strategy for community-defined and community-driven open surveys, producing fully public data, while retaining flexibility for general astrophysics investigations. Thus, Roman's WFI observing portfolio will include large, pre-defined Core Community Surveys (CCS) and a limited set of General Astrophysics Surveys (GAS), defined by a community-led process (\autoref{app:world_contributions}) and traditional peer-reviewed calls for proposals, respectively. 

Roman's Core Community Surveys comprise three large projects: the High Latitude Wide Area Survey (HLWAS; \autoref{sec:hlwas}), the High Latitude Time Domain Survey (HLTDS; \autoref{sec:hltds}), and the Galactic Bulge Time Domain Survey (GBTDS; \autoref{sec:gbtds}). The effort of crafting observational strategies for each CCS was shepherded by a Definition Committee (DC); each DC designed a nominal, descoped (underguide), and enhanced (overguide) survey strategy to meet Roman's science requirements and enable additional science. The Roman Observations Time Allocation Committee (ROTAC) was charged with reviewing these observational strategies and making time allocation recommendations for the CCS and the GAS programs. In addition, the Coronagraph Participation Group will execute an observing program with Roman's Coronagraph Instrument using 3~months of observing time during the first 1.5~years after launch; while this is formally a technology demonstration program, we expect that it will have substantial science return as well. The ROTAC charge, and thus the contents of this report, are limited to the WFI observations.

\begin{figure}[!h]
\includegraphics[width=\textwidth]{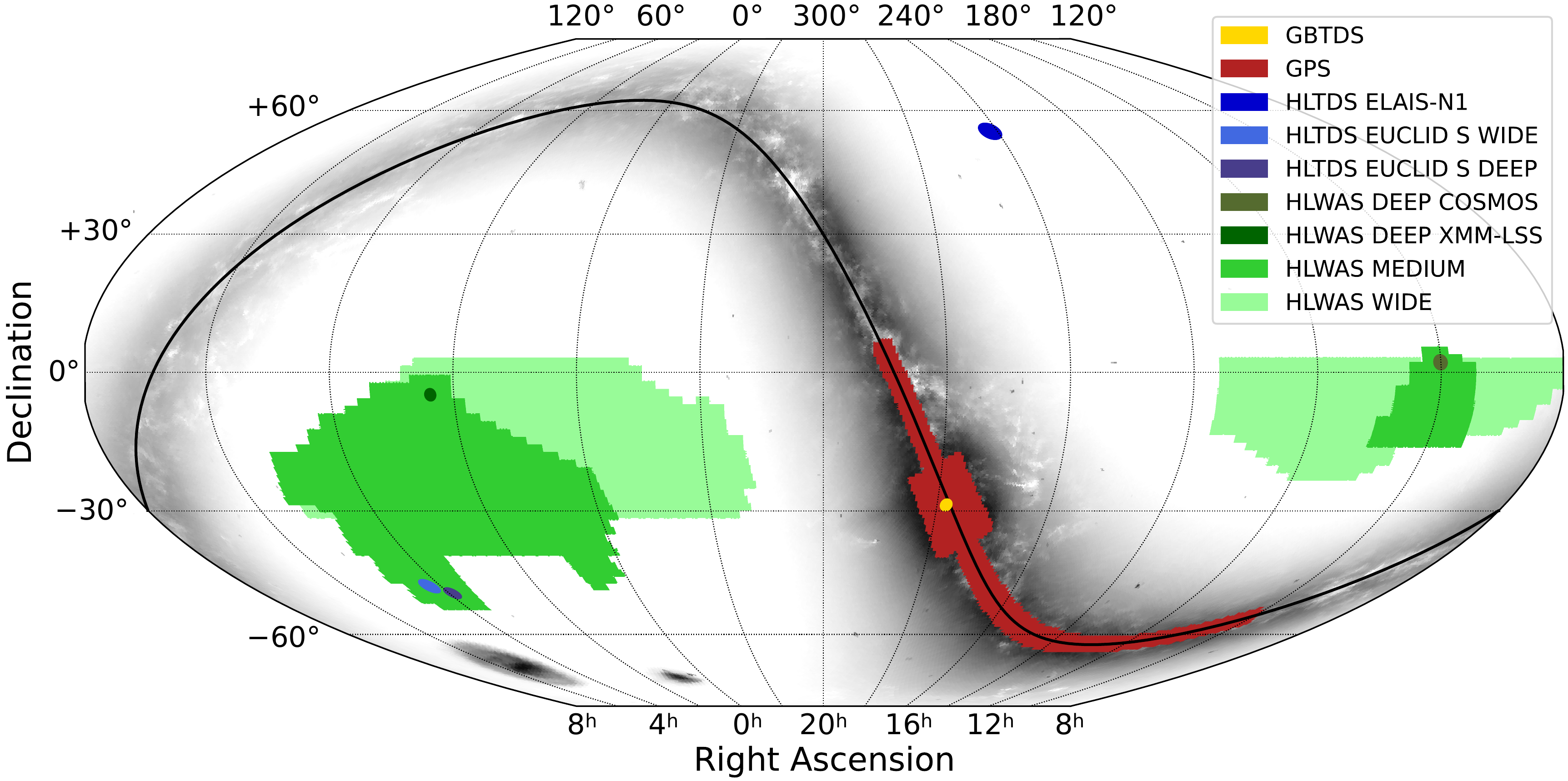}
\caption{Equatorial projection of the current footprint of the recommended CCS design options: HLWAS tiers in light green (wide tier), medium green (medium tier), and dark green (deep tiers); HLTDS in blue; and GBTDS in yellow (see \autoref{fig:gbtds_field_layout} for a more detailed GBTDS field layout). Also shown in red is the early-definition Galactic Plane Survey (GPS). The total footprint is here overlaid on a background stellar density from \emph{Gaia}. Figure code credit: Javier~Sanchez and Saurabh W.~Jha, with assistance from Eli Rykoff and Alex Drlica-Wagner; data credit: ESA/Gaia/DPAC.}
\label{fig:ccs_skymap}
\end{figure}

\begin{figure}[!h]
\includegraphics[width=\textwidth]{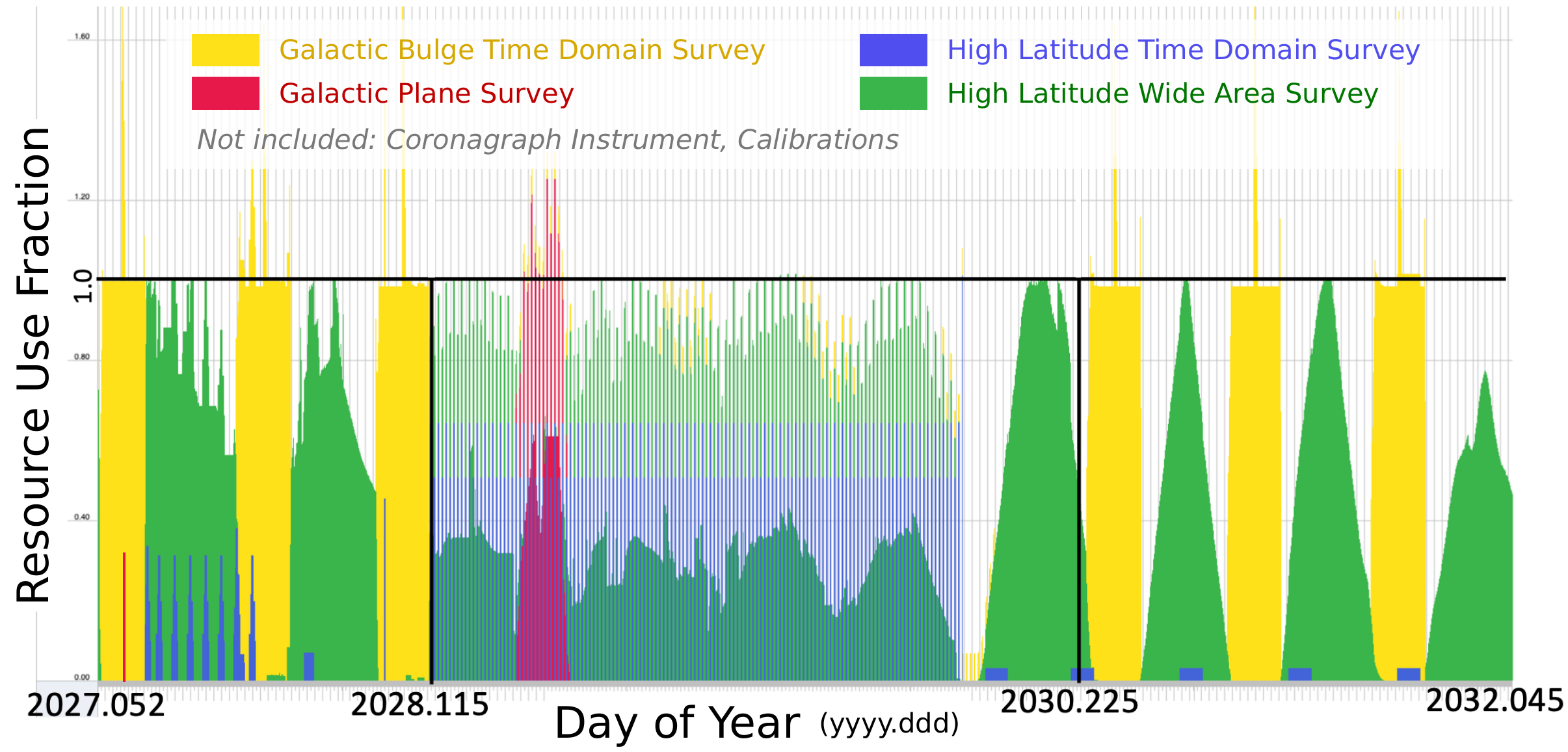}
\vspace{-1cm}
\caption{Illustrative simulation of the Roman observing timeline. The GBTDS observes nearly continuously in the first three Galactic bulge seasons and then again late in the mission. The HLTDS main survey observes at high cadence during the middle years of the prime mission. The HLWAS and GPS are scheduled around the time-domain surveys, but all of the programs obtain data early on in the mission. This timeline will be updated to include observations with the Coronagraph Instrument and Calibrations, and it will shift depending on the actual start date of survey operations (a late 2026 launch was assumed here). The brief periods when resource demand exceeds availability ($>$1.0) are largely artifacts here and will be rectified. The remaining available observing time will be dedicated to General Astrophysics Surveys. Figure credit: Roman Science Operations Center at STScI.}
\label{fig:observing_schedule}
\end{figure}

The footprint of our recommended observing plan is shown in \autoref{fig:ccs_skymap}, with an illustrative simulated observing timeline in \autoref{fig:observing_schedule} that assumes a realistic launch date of October 2026\footnote{\url{https://asd.gsfc.nasa.gov/roman/comm_forum/}}. Below we give a short summary of each CCS and in \autoref{sec:ccs}, our reasoning behind our recommendation for each design plan. We refer the interested reader to the fuller reports from each survey DC (\autoref{app:dc_reports}),
upon which these summaries are based and which contain far more detail. {\bf The ROTAC is extremely grateful to the DCs for their thoughtful and extensive work on these surveys and their reports.}

The Galactic Bulge Time Domain Survey (GBTDS) will observe five 0.28 deg$^2$
fields in the Galactic Bulge and a sixth in the Galactic Center. GBTDS observing
will be concentrated in six 72-day seasons, three early in the mission and
three late in the mission.  During these seasons, each of the six fields 
is observed every 12.1 minutes, in the wide W (F146) filter, with an exposure 
time of 67 seconds, and every 12 hours or faster in the Z (F087) and K (F213) filters, with final filter and cadence choices to be determined through community input. In the four
intervening seasons, fields will be observed every 3--5 days to provide coverage of long timescale events. The GBTDS
will also obtain multiband photometry and grism snapshots of each field for characterization of its stellar populations.
The GBTDS is designed to measure the demographics of exoplanets via 
gravitational microlensing, sensitive to a regime of mass and orbital
separation that is complementary to those of the Kepler and TESS missions,
including free floating planets. The GBTDS will also detect stars and
black holes via microlensing, detect more than $10^5$ transiting planets, measure
asteroseismic masses for over $10^5$ evolved stars, and enable a wide range
of other studies of stellar and Galactic astrophysics.

The High-Latitude Time Domain Survey (HLTDS) will carry out imaging and
spectroscopy with an approximately 5-day cadence over a 2-year duration in
the middle of Roman's 5-year mission. Imaging will cover a wide tier area
of about 18 deg$^2$ in RZYJH (F062, F087, F106, F129, F158) filters and a deep tier area of about 6.5
deg$^2$ in ZYJHF (F087, F106, F129, F158, F184) filters, with each tier having both northern and southern
fields.  Exposure times, ranging from $\sim$90--300 sec in the wide tier
and $\sim$300--1600 sec in the deep tier depending on band, are designed to
reach S/N $\sim 20$ for a typical type-Ia supernova (SN Ia) at maximum light at redshifts $z \sim 0.9$
(wide) or $z \sim 1.7$ (deep).  The prism spectroscopic observations will cover
4.5 deg$^2$ and 0.56 deg$^2$ in the wide and deep tiers, respectively, 
providing redshifts and classifications of a subsample of transients.
The HLTDS will also include a pilot component early in the mission
to provide templates for difference imaging and data for calibrating methodology 
and measuring supernova rates. The HLTDS is designed
to probe the origin of cosmic acceleration by using SN Ia to make
precision measurements of the cosmic expansion history.  It will also 
enable a wide range of extragalactic time-domain and transient science 
investigations and provide deep co-added imaging and spectroscopy over
areas 1--2 orders of magnitude larger than any comparably deep Hubble imaging.

The High-Latitude Wide Area Survey (HLWAS) will observe a 2400 deg$^2$ 
medium tier in 
YJH (F106, F129, F158) imaging and grism spectroscopy, with characteristic imaging depth of
26.5 mag (AB, $5\sigma$ point source, $t_{\rm exp}\approx 600$ sec) 
and spectroscopic depth of of $1.5\times 10^{-16}$ erg/cm$^2$/sec
($5\sigma$ line flux limit, $t_{\rm exp} \approx 1500$ sec).
A wide tier will 
cover an additional 2700 deg$^2$ in H-band (F158) imaging only, with the same
exposure time.  A deep tier of 19.2 deg$^2$ with longer cumulative 
exposure time will reach about 1.2~mag deeper in imaging and a factor of 2.5
deeper in line flux, and it will include Z (F087), F (F184), K (F213), and wide W (F146) filters
as well as YJH. A 5~deg$^2$ subset of this tier will have additional
YJH imaging to reach 0.5~mag further depth in these three bands. The HLWAS imaging survey
is designed to probe the origin of cosmic acceleration by using weak 
gravitational lensing and galaxy clustering to make precision measurements
of cosmic expansion and the growth of cosmic structure.  The HLWAS 
spectroscopic survey is designed to achieve similar goals using precision
measurements of baryon acoustic oscillations (BAO) and redshift-space distortions
in galaxy clustering.  These data will also enable a wide range of science
investigations including solar system objects, Galactic structure, nearby 
galaxies, and galaxy and quasar evolution out to the epoch of reionization.
The deep tier provides necessary calibration data for the medium and wide
tiers, and it will be a powerful resource for studying faint stellar 
populations and the most distant galaxies and quasars.  HLWAS observations
will be spread throughout the 5-year mission, with at least half of the
deep tier observations carried out early in the mission and most of the
wide tier observations carried out late in the mission.

While the Core Community Surveys will already support science goals far beyond 
those used to set their technical requirements, the General Astrophysics Surveys
program will enable investigations tailored to specific science goals.  With up to 30
programs allowed in the 389 day allocation, the mean observing time for a 
General Astrophysics Survey is about 13 days, or 1.1 Msec.  Based on
the recommendations of Roman's Early Science Definition Committee, 
a General Astrophysics Survey of the Galactic plane is being defined now.
While details are still to be determined, this one-month survey will likely 
image several hundred deg$^2$ of the Galactic plane in one or two filters and
survey a much larger area of the Galactic bulge than the GBTDS (\autoref{fig:ccs_skymap}).
The Galactic plane survey is planned to be executed during the first two years 
of the mission, nominally during the first season when the GBTDS is not observing at high cadence (\autoref{fig:observing_schedule}).  
While some other GAS time will also be allocated during
the first two years, most of the GAS observing will occur later in the 
mission, when the completion of some programs makes scheduling more flexible,
and when investigations using the early Roman data have given the community
more complete knowledge of the observatory's capabilities.
GAS programs will be selected through a combination of open, competitive proposals 
and community-defined processes.

All Roman data, from both Core Community Surveys and GAS programs, will be made
public through the Science Operations Center (SOC) at STScI as soon as the basic
data reduction is complete.  There is no proprietary period on Roman data.

The ROTAC recommendations cover the 5-year prime mission of the Roman Space
Telescope. The observatory is designed to accommodate a much longer lifetime, so we
anticipate that even the ambitious program described here represents just the
first phase of Roman's long-term scientific contributions, discoveries, and impact.

\section{Implementation of the Core Community Surveys}
\label{sec:ccs}

\subsection{High Latitude Wide Area Survey (HLWAS)}
\label{sec:hlwas}

The goals of the HLWAS are to probe the growth of cosmic structure and expansion of the Universe via galaxy clustering and gravitational lensing, while also enabling a wide range of general astrophysics via a deep, wide-field infrared survey. Key design considerations included the balance of imaging at different depths and spectroscopy for this wide-area high-impact survey, the field placement with respect to existing complementary data (including Euclid), and the detailed selection of filters.

The ROTAC recommends adoption of the nominal (in-guide) survey proposed by the HLWAS DC. The core of the survey is a ``medium tier'' of YJH imaging and grism spectroscopy over 2415~deg$^2$.  A ``deep tier'' of 19.2~deg$^2$ provides essential calibration data and will also be a powerful survey of faint objects.  A ``wide tier'' expands the HLWAS imaging area by an additional 2700~deg$^2$ in H-band only.

These three tiers enable precise cosmological measurements that meet the mission requirements. Imaging in the wide and medium tiers will be utilized for weak lensing measurements of the growth of structure, with multi-band observations in the medium tier providing excellent control of systematics. The grism spectroscopy in the medium tier will enable geometric measurements of cosmic expansion via BAO and large-scale structure constraints via redshift space distortions. The deep tier (including ultra-deep imaging) will provide critical calibration of the cosmological analyses including shape measurements, photometric redshifts, and spectroscopic decontamination. 

Each of the three tiers also offers distinct value in meeting the broader mission requirement of deep, wide general astrophysics surveys. 

\vspace{-12pt}
\begin{description}
    \item[Medium Tier:] The medium tier, which comprises YJH imaging and grism spectroscopy over 2415~deg$^2$, has a characteristic depth of 26.5~mag (AB) in imaging ($5\sigma$ point source) and $1.5 \times 10^{-16}$ erg/cm$^2$/sec in spectroscopy ($5\sigma$ emission line integrated flux limit).  It is expected to yield about 360 million galaxy shape measurements for weak lensing (with $n_{\rm eff} \approx 41$ arcmin$^{-2}$) and 19~million spectroscopic galaxy redshifts. 

    The multi-band, Hubble-resolution imaging and deep spectroscopy will also enable the study of local galaxies, Milky Way stellar populations, and the relation between galaxies and dark matter over a vast span of redshift. 
    
    \item[Deep Tier:] The deep tier covers two of Rubin's near-equatorial deep drilling fields, COSMOS and XMM-LSS, 
in ZYJHFK, the wide W band, {\it and} in grism spectroscopy. In YJH, the imaging will be approximately 1.2~mag deeper than in the medium tier, and the same exposure times will be used for the other filters, with $5\sigma$ point source depth ranging from 25.9 AB mag (K, F213) to 28.3 AB mag (W, F146).  The total spectroscopic exposures will be 8$\times$ longer than those in the medium tier, with a line flux limit about 2.5$\times$ deeper.  The deep tier also includes an ultra-deep component, imaging 5~deg$^2$ in YJH to a depth 1.7~mag fainter than the medium tier, for calibrating noise biases in weak lensing shape measurements.  
 
    The deep tier will also enable studies of faint objects including the most distant galaxies, with depth $\sim$0.5 mag deeper at near-IR wavelengths than the HST~CANDELS program, in more bands, over an area 86$\times$ larger.  The ultra-deep component will probe another 0.5 mag deeper over 1/4 of this area. The deep tier spectroscopy will probe the physical properties of distant and faint galaxies, reaching a similar flux limit to 3D-HST over an area 100 times larger. The deep tier also includes dense time sampling that will make it valuable for solar system science and some time-domain applications. 
    
    \item[Wide Tier:] The wide tier uses 80~days (15\% of the total HLWAS allocation) to more than double the total survey area, albeit in a single band (H, F158). The exposure time matches that of the medium tier, but the characteristic depth is slightly shallower (26.2 mag AB) because of a higher zodiacal background in the available fields. The wide tier is expected to yield an additional 260~million shape measurements ($n_{\rm eff} \approx 27$ arcmin$^{-2}$).

    The wide and medium tiers together yield over 5000 deg$^2$ of high-latitude sky with Hubble-like near-IR imaging, all of which should also have optical photometry from Rubin. The gains from the wide tier are still larger when considering overlap with other wide-field multiwavelength surveys. For example, the wide tier greatly expands the overlap with the DESI spectroscopic galaxy survey.
\end{description}

For the underguide and overguide scenarios (detailed descriptions in the DC report; \autoref{app:hlwas}), the HLWAS~DC maintained the structure of the medium and deep tiers and focused on varying the area covered by the wide (H-band imaging only) tier. The consensus of the ROTAC was that the underguide scenario, which entailed reductions to the HLWAS wide tier in both the north and south Galactic cap regions, would require too much loss of science that is distinctively achievable with Roman. Conversely, the ROTAC felt that the overguide scenario, which focused on expanding the wide tier area in the South Celestial Pole (in part to maximize overlap with the South Pole Telescope),
did not offer sufficiently broad impact to require inclusion in the CCS design.

One topic of discussion for the ROTAC when considering the HLWAS was the unique capabilities of Roman for the wide tier compared to the ongoing wide-area survey being carried out by the ESA Euclid mission. Euclid's NIR filters provide lower sensitivity and resolution than Roman at those wavelengths, so the most relevant comparison is between the HLWAS wide tier and the Euclid visible (VIS) survey. The discussion highlighted a number of key distinctions: Roman data are immediately available to all (and the US\,community in particular), unlike Euclid data; the HLWAS wide tier imaging utilizes a standard NIR band (H) versus Euclid's wide, non-standard optical VIS band that is also more vulnerable to chromatic PSF effects that may bias weak lensing measurements; and Roman's H-band has moderately higher angular resolution ($\approx$0.1$^{\prime\prime}$) than Euclid VIS (0.16$^{\prime\prime}$). In addition, the HLWAS wide tier fills out the equatorial band that will have Rubin LSST coverage and is accessible from the Northern Hemisphere, including DESI, while Euclid's primary survey (while observing much more of the sky overall) covers only $\sim$470 deg$^2$ of the 2700 deg$^2$ HLWAS wide tier. These distinct capabilities justified adopting the full in-guide area of the HLWAS wide tier.

In concurrence with the HLWAS DC, the ROTAC recommends scheduling at least half of the deep tier observations early in the mission, both for calibration and for galaxy and quasar evolution science. 
The wide tier, on the other hand, is better scheduled for the later phases of the mission.

\subsection{High Latitude Time Domain Survey (HLTDS)}
\label{sec:hltds}

The primary goal of the HLTDS is to discover and follow up SNe~Ia to measure the expansion history of the Universe. 
The HLTDS will also potentially discover exotic high-redshift transients, dust-extinguished sources, and cool/red objects. Furthermore, the NIR transient sky has been less well explored to date, offering the opportunity for serendipitous discovery.

The primary survey design considerations included filter combination, depth and cadence, value of spectroscopy, field selection, and survey extensions.  The HLTDS Committee recommended a survey that takes advantage of Roman's unique capabilities to target SN~Ia at redshifts not easily accessible from the ground.  Simulations showed that at least two tiers are required to cover a broad redshift range from $z = 0.5 - 2.5$, with observations in four or five filters to measure accurate colors in the rest-frame optical and NIR.  Exposure times and cadences were chosen to accumulate a fixed (on average) S/N at the target redshifts.

The HLTDS DC restricted the survey fields to lie in the continuous viewing zones (CVZ) for continuous visibility through the year.  Multiple fields are desirable for spectroscopic follow-up observations on multiple ground-based telescopes and to also reduce cosmic variance.

Due to constraints on the telescope's roll angles, the mosaic pattern of the fields must be roughly circular to reduce the amount of edge effects as the fields are observed through the year.  This is crucial for reducing areas with poor cadence especially for mosaics with small number of pointings.  Thus, the feasible areas are quantized.

\begin{figure}[H]
\begin{center}
\includegraphics[width=0.75\textwidth]{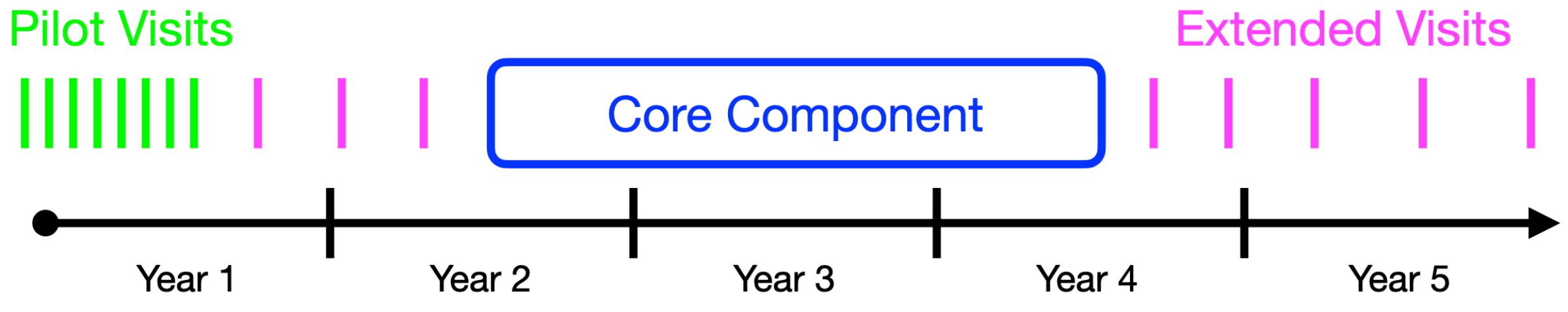}
\includegraphics[width=0.75\textwidth]{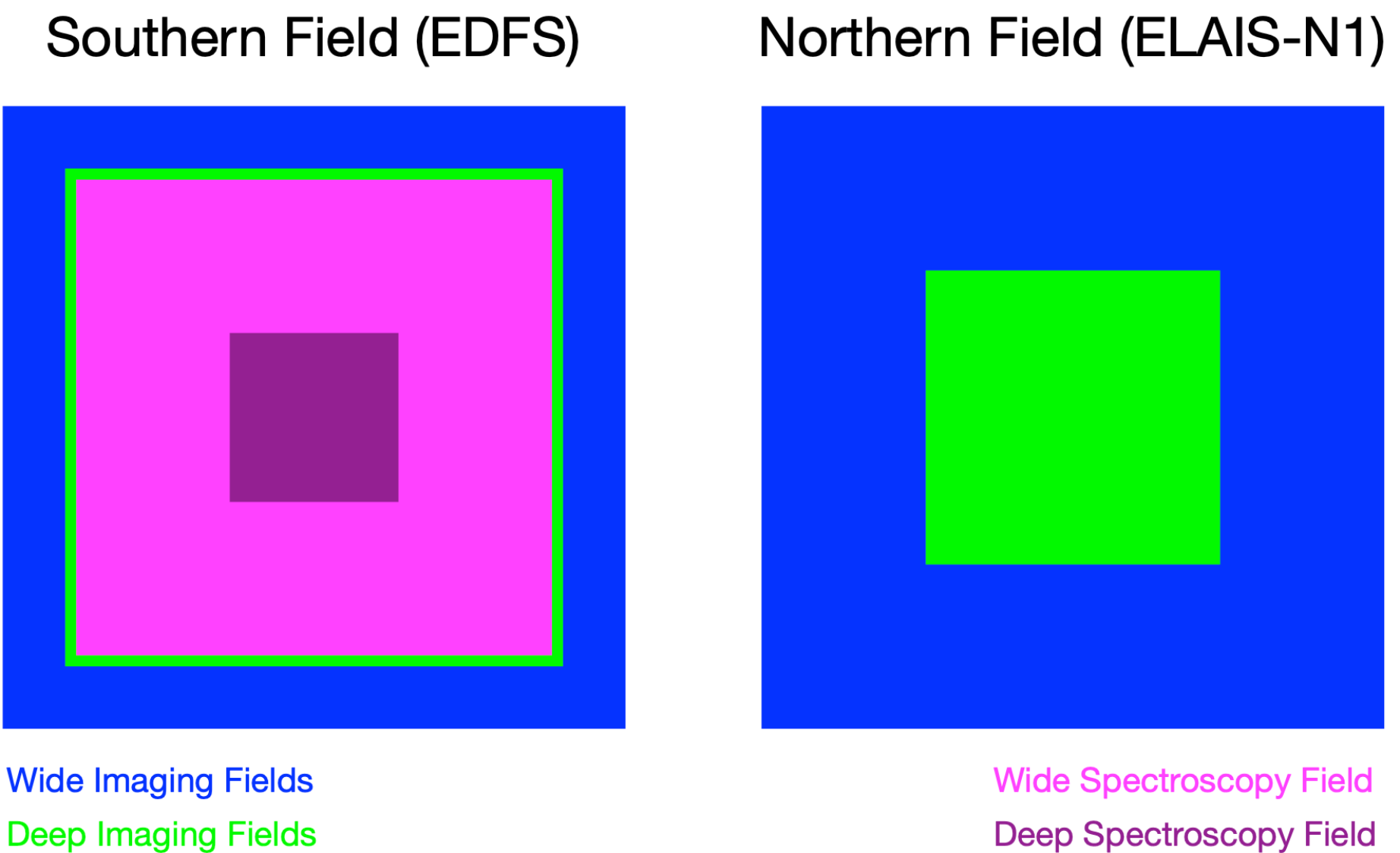}
\vspace{-0.9cm}
\caption{Top: nominal timeline of HLTDS components. Bottom: Relative area covered by the different imaging and spectroscopy tiers. Graphic from the HLTDS DC report (\autoref{app:hltds}).}
\label{fig:hltds}
\end{center}
\end{figure}

The nominal/in-guide plan, which is the one recommended by the ROTAC, comprises the following three components:
\vspace{-12pt}
\begin{description}

    \item[Core Component:] The heart of the HLTDS will take place over a 2-year duration approximately in the middle of Roman's 5-year mission.  It consists of two fields --- coincident with the ELAIS-N1 field in the North ($\alpha,\delta=242.504, +54.510$ deg) and the Euclid Deep Field South (EDFS) in the South ($\alpha,\delta=61.241, -48.423$ deg) --- layered with tiers of both imaging and spectroscopy. The Wide Imaging Tier will cover 10.68~deg$^2$ in the North and 7.59~deg$^2$ in the South with RZYJH filters, with a $\sim$10-day cadence of alternating filters (i.e., one sequence of RZY or RJH every $\sim$5 days), to reach an average maximum-light S/N of 20 for SN~Ia at $z \sim 0.9$. The Deep Imaging Tier will cover 1.97~deg$^2$ in the North and 4.5~deg$^2$ in the South with ZYJHF filters, with a similarly interlaced sequence of ZYJ and ZHF images, to reach an average maximum-light S/N of 20 for SN~Ia at $z \sim 1.7$.  The expected number of ``good'' SNe~Ia, defined to have a cumulative S/N $>40$, is $\sim 7500$ and $\sim 6800$ in the Wide and Deep Tiers, respectively.
    
    The Wide and Deep Spectroscopy Tiers, covering 4.5~deg$^2$ and 0.56~deg$^2$, respectively, will both take place in the deep imaging area in the South, with a $\sim$5~day cadence.  The exposure times in the Wide and Deep Tiers are 900 sec and 3600 sec, respectively, providing redshifts and classification of a subsample of transients in that field. The total allocation of this component is 158 days.
    
    \item[Pilot Component:] This component will take place before the Core Component and as early as possible in the mission.  It serves multiple purposes -- 1) collection of template images for difference imaging, 2) collection of reference prism data for host galaxy subtraction/modeling and 3) measurement of the rate of SNe~Ia above $z \sim 1$.  A series of 8 observations will be made on a cadence of 20 days, thus enabling light curve measurements and data acquisition over multiple roll angles.  The total allocation of this component is 15 days.
    
    \item[Extended Component:] This component extends the temporal coverage of the Deep Imaging Tier over Roman's entire 5-year mission.  A total of 8 observations will be made in between the Pilot and Core Components as well as after the Core Component with an approximate cadence of 120 days.  This will enable discoveries of the highest-redshift super-luminous and pair-instability SNe possibly out to $z \sim 5$.  The total allocation of this component is 7 days.
    
\end{description} 

Detailed descriptions of overguide and underguide survey designs are given in the DC report (\autoref{app:hltds}).
The underguide design option, which takes 20 days less than the in-guide, would reduce the area of the Wide Imaging Tier by nearly 20\% --- from 18~deg$^2$ to 14.5~deg$^2$ --- without any other adjustments.  This would result in about 1500 fewer SN~Ia. The ROTAC does not recommend this option.

The overguide design, which extends 20~days longer than the recommended nominal design, consisted of additional imaging observations in the Deep Tier in the K and R filters to complete the full filter set. This would enhance galaxy science and improve photometric redshift measurements. The ROTAC does not recommend this option, as the enhancement to the core goals of the program are incremental.

\subsection{Galactic Bulge Time Domain Survey (GBTDS)}
\label{sec:gbtds}

The primary science driver for the GBTDS is the demographics of small, long-period exoplanets. This science will be enabled by obtaining high cadence ($<$15~minute), high precision time series photometry over a $\sim$2-square degree field of view towards the Galactic bulge, in order to observe microlensing events caused by both bound and free-floating planets. 

As with the other two CCS above, the field placement and filter selection were important design considerations, but the GBTDS involved a particularly complex consideration of tradeoffs between survey area and imaging cadence. Each design option included a region of 5--7 contiguous fields at $(\ell \approx -0.5^\circ$, $b \approx -1.2^{\circ})$, which was selected to optimize stellar density against the impacts of dust extinction. They also all included a field on the Galactic center, which was a field of particular interest to the general astrophysics community and was recommended by many white papers.  

\begin{figure}[H]
\begin{center}
\includegraphics[width=0.7\textwidth]{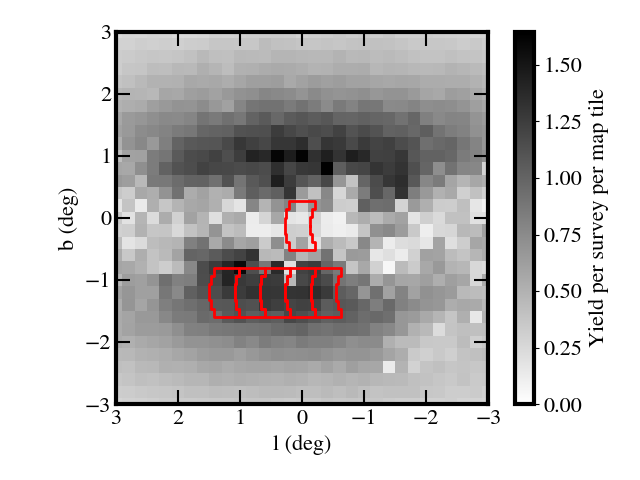}
\vspace{-1cm}
\caption{
Field layout for the overguide GBTDS survey as recommended by the ROTAC. Background contours show the expected yield for  1~$M_{\rm Earth}$ free-floating planets. Graphic from the GBTDS DC report (\autoref{app:gbtds}).
}
\label{fig:gbtds_field_layout}
\end{center}
\end{figure}

The ROTAC recommends adoption of the overguide survey outlined by the GBTDS DC, with some modifications as detailed below. \autoref{fig:gbtds_field_layout} shows the current field layout of this design. This survey comprises four main components:
\vspace{-12pt}
\begin{description}
    \item[High-Cadence Seasons:] Observations of the survey fields in six of the 10 available bulge seasons, with a cadence of 12.1~minutes with the wide W (F146) filter and a cadence of 12~hours or faster with the Z (F087) and K (F213), or other recommended filters to achieve the science requirements of the exoplanet microlensing survey. These seasons should be scheduled as the first three and the last three bulge seasons over the 5-year duration of the Roman mission in order to maximize the astrometric baseline, and observations are recommended to span the full duration of each bulge season to maximize the photometric baselines. 
    
    \item[Low-Cadence Seasons:] Observations of the same high-cadence survey fields over the four remaining bulge seasons with a cadence of three days with the W (F146) filter and the K (F213) filter (but see below) to enable the recovery of long-duration microlensing events and the detection of multiple types of variables. It is recommended to take four exposures per visit, for stacking to improve photometric and astrometric precision and cosmic-ray rejection, and to optimize the science time-to-slew time ratio. Field locations are recommended to be identical to the high-cadence seasons in order to preserve the target sample, photometric and astrometric baselines.
    
    \item[Multiband Photometry Snapshots:] Observations of the survey fields with the five filters that are not used in the high-cadence or low-cadence seasons for stellar characterization. These observations will take place at the start, middle, and end of each of the ten bulge seasons, for a total of 30 snapshots over the survey.
    
    \item[Spectroscopic Snapshots:] Observations of the survey fields with the grism to enable measurements of stellar temperatures, metallicities and radial velocities. As with the multiband photometry, these observations will take place at the start, middle, and end of each of the ten bulge seasons, for a total of 30 epochs.
\end{description}

Detailed descriptions of the nominal, overguide, and underguide survey designs are given in the DC report (\autoref{app:gbtds}).
In short, the underguide design included shorter total observing time per bulge season, longer cadences in both the high- and low-cadence seasons, much longer cadence in the Galactic Center field, and fewer snapshots, but in a larger number of fields (in order to meet mission requirements with the complex interplay between survey area and duration). The ROTAC does not recommend this option.

As this CCS is the only one for which ROTAC is recommending the larger overguide time allocation over the nominal allocation, we describe our decision-making process here in greater detail than in the subsections above.

The nominal design is, unsurprisingly, intermediate between the underguide and overguide designs in terms of per-season observing allocation, number of fields, cadence, and number of multiband$+$grism snapshots. The ROTAC solicited further simulations from the GBTDS DC for some of the ancillary science cases (such as transiting exoplanets and asteroseismology) using the current field layouts and survey parameters, since such simulations were not available in time for the DC report. As a result, in addition to the underguide, nominal, and overguide designs, the ROTAC considered a survey with the nominal time allotment (420 days), but with 5 contiguous fields and a 12.1~min cadence.

\begin{figure}[ht!]
\begin{center}
\includegraphics[width=\textwidth]{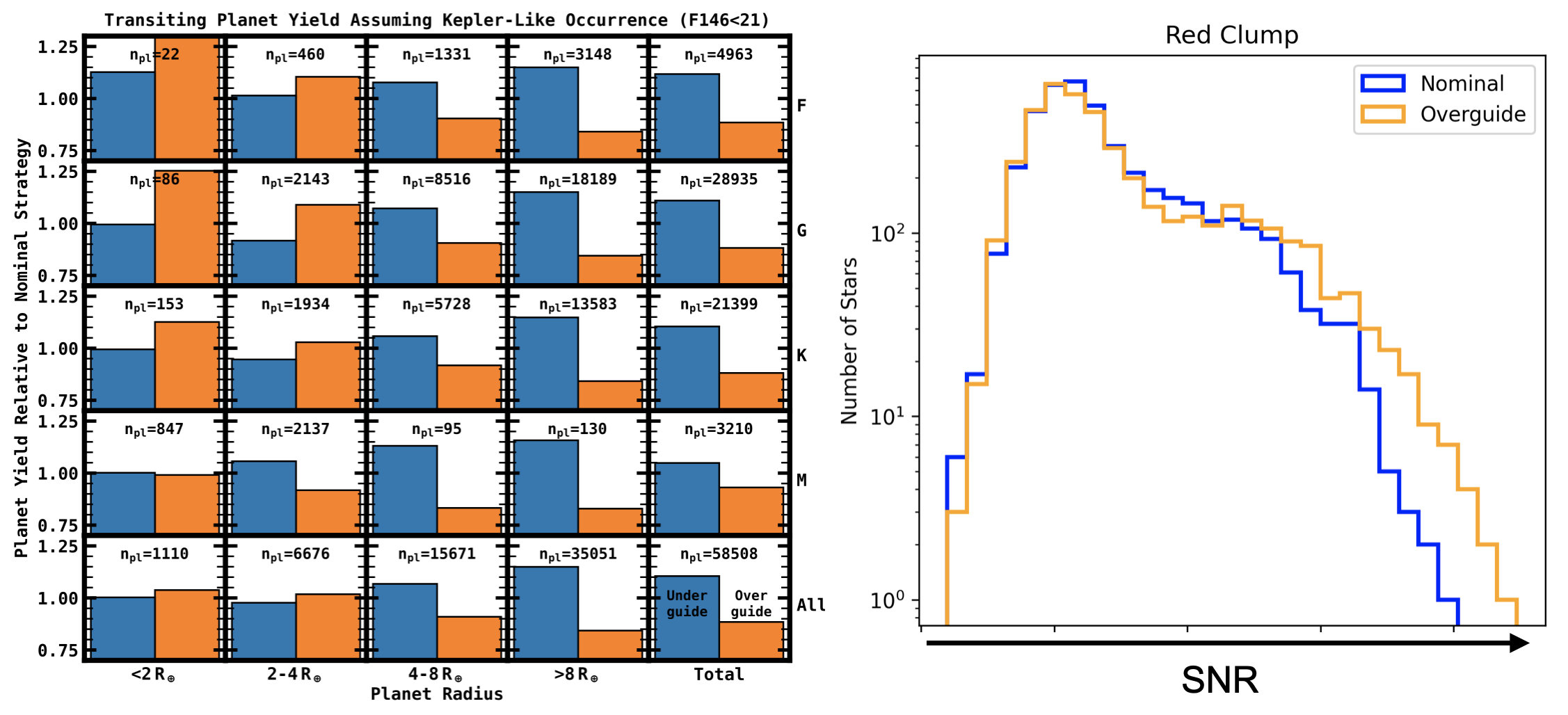}
\vskip -1cm \caption{Yield simulations for transiting planets and asteroseismology using the recommended survey designs. Left: Transiting exoplanet yield relative to the nominal survey as a function of planet radius (x-axis) and host star spectral types (y-axis). Blue bars denote the underguide, and orange bars the overguide survey. Numbers in each panel denote the number of planets expected to be detected in the nominal survey design. Figure credit Robby Wilson (GSFC). 
Right: Signal-to-noise ratio histogram for asteroseismic detections in red clump stars for the nominal and overguide recommendations. 
Figure credit: Trevor Weiss (CSULB) and Noah Downing (OSU).
}
\label{fig:gbtds_yield}
\end{center}
\end{figure}

In deciding between the survey designs, the ROTAC was persuaded as to the value of higher cadence observations (12.1~min rather than~14.8 min), even at the price of a smaller survey area. Higher cadence observations are expected to yield better sensitivity to small transiting planets, which are a particular strength of Roman at a large range of Galactic distances,
though the decreased number of fields leads to fewer detected transits by large planets (left panel of \autoref{fig:gbtds_yield}).
While a higher-cadence/fewer-fields survey design leads to fewer asteroseismic detections (a $\sim$7\% decrease in the overguide compared to the nominal), the resulting detections will have higher S/N and thus enable more precise measurements (right panel of \autoref{fig:gbtds_yield}), which both the GBTDS DC and ROTAC deemed more compelling than high numbers of noisy detections.   

While microlensing yields are indeed smaller in a higher-cadence/fewer-fields survey design (see Tables~5 and 6 of the GBTDS DC report; \autoref{app:gbtds}), the expected decrement is very modest and did not significantly concern the ROTAC. A higher-cadence survey is expected to detect $\sim$5\% more $\lesssim$M$_{\rm Mars}$ free-floating planets, and $\sim$10\% fewer planets with $\gtrsim$10 M$_{\Earth}$, compared to the nominal design; ROTAC decided that it was compelling to prioritize the lower mass planets. The higher-cadence survey will also yield $\sim$15\% more finite source measurements, which will translate to more mass measurements for planet microlensing events. While these simulations (some of which were made after the DC report, at the request of the ROTAC) highlight that there are challenging trade-offs between the 14.8 min/6-contiguous-fields nominal design and a 12.1 min/5-contiguous-fields scenario, the ROTAC ultimately felt that 14.8~min cadence skirted the edge of what is necessary for important GBTDS science drivers, and furthermore that a 12.1~min cadence leaves a more comfortable margin for extracting low-mass planets (see also \autoref{sec:uncertainty} below), high-quality asteroseismic signals, and serendipitous events from the data. The ROTAC was concerned that critical information could be lost and made unrecoverable, should the lower cadence design be implemented. The committee therefore converged on a recommendation for 5 contiguous fields plus a Galactic center field at a 12.1~min cadence.

In deciding between the nominal and overguide time allotments, the ROTAC explicitly considered each of the following components: the additional 12~days that would be allotted to monitoring during the high-cadence seasons, the additional 3~days that would be allotted to monitoring during the low-cadence seasons, the additional 3~days that would be spent on multi-band photometric snapshots, and the additional 2~days that would go to grism snapshots. The ROTAC was compelled by the fact that the overguide design prioritizes broader astrophysics questions (e.g., black hole microlensing, Sgr A*, and binary stars) at the price of slightly diminished yields for the core science of planet microlensing. However, the overguide will still comfortably meet the Level~1 science requirements, under the assumption of the \citet{Cassan2012} mass function (but see \autoref{sec:uncertainty} below).

The ROTAC was convinced of the value of spending as much time on the GBTDS fields as possible during the high-cadence seasons, and generally supported the increase of 2~days per season suggested by the overguide design. The longer seasons will partially compensate for the smaller survey area necessitated by the faster 12.1-min cadence and also for the lower microlensing planet yields of the updated \citet{Suzuki2016} mass function.
However, during the first Galactic bulge season, there is a need for wider field observations as part of the Early-Definition Galactic Plane Survey\footnote{\url{https://roman.gsfc.nasa.gov/science/galactic_plane_survey_definition.html}}, to enable proper motion measurements of stars in the that survey's footprint. Therefore, the ROTAC recommends that GBTDS be allotted 70.5~days in the high-cadence bulge seasons, but that the first bulge season should be reduced by an amount that will enable the Galactic Plane Survey to conduct its first epoch of observations (anticipated to be $\sim$2~days).

The ROTAC was persuaded of the value of additional monitoring observations in the low-cadence seasons, as moving from a 5~day to a 3~day cadence is demonstrated to increase the number of black hole microlensing mass measurements at fixed precision by $\approx$\,10\%
(Figure~9 of the GBTDS DC report; \autoref{app:gbtds}). However, the committee considered how scheduling these 3-day-cadence observations would interweave with the 5-day cadence of the HLTDS (which requires large slews between the north and south). The ROTAC was concerned that the observing cadence of the two surveys could beat against one another in inefficient ways, potentially leading to large overheads and limiting flexibility of Galactic bulge or other GAS science. It is likely that similar scientific benefits of the GBTDS low-cadence season observations could be reached either with longer duration visits on the nominal 5~day cadence or with flexible, aperiodic cadence that seeks to approximate a $\sim$3~day cadence, while maintaining efficient observatory operations. Therefore, the ROTAC's recommendation for this component of GBTDS is to increase the time allotted to the low-cadence season monitoring by 3~days above nominal, but 
for the Roman project to work closely with the relevant scientific and technical experts
to best schedule this additional time.

The ROTAC was also persuaded of the broad scientific value of the additional photometric and grism snapshots, which significantly improve the window function for time series analyses. The confusion-limited sensitivity of $K \lesssim 16$~mag will yield grism spectra for hundreds of thousands of stars and $\sim$10 km~s$^{-1}$ radial velocity precision, while photometric timeseries with all of the Roman bands will provide additional insights into a range of science questions related to stellar variability. The ROTAC therefore recommends the GBTDS be allocated the additional 2~days for these observations.

Thus, the ROTAC generally endorses the overguide survey design for GBTDS, with two important caveats: that the first high-cadence bulge season be modestly reduced to leave time for the Galactic Plane Survey, and that the observation schedule during the low-cadence seasons be further studied to minimize disruption to the HLTDS and GAS. 

\section{General Astrophysics Surveys}
\label{sec:gas}

The ROTAC placed a high priority on enabling the community to propose general astrophysics surveys (GAS) that will address compelling science goals beyond those that can be accomplished by the CCS, and we adopted a stewardship role towards the valuable GAS observing time. 
As with the CCS, the GA programs will be limited to the wide-field instrument. 

The prime mission profile recommended by the ROTAC allocates 389~days for GAS, after accounting for the recommended allowances for the HLWAS, HTLDS, and GBTDS programs. This GAS allocation is well above the minimum requirement that $\ge$15\% of Roman's prime mission science time be reserved for such work; indeed it slightly exceeds the mission {\it goal} of making 25\% of the time available.

This time is currently planned to be allocated to a maximum of 30 approved WFI GA surveys that will be solicited across a subset of Roman calls for proposals (alongside proposals for analysis resources only). 
This subset is expected to span 2026--2029, with most of the GAS time being allocated in the later cycles when the observatory performance is more fully understood. The ROTAC advocates that the community organize to generate proposals for relatively large surveys, designing each to address broad science cases, to take better advantage of Roman's capabilities than many different small programs would. As an example, one such GAS program has already been approved by the Early-Definition Astrophysics Survey Assessment Committee: a survey of the Galactic Plane\footnote{\url{https://roman.gsfc.nasa.gov/science/galactic_plane_survey_definition.html}} that will take 30 days of the GAS observing time.
Moreover, numerous science cases recommended to the CCS DCs through white papers were judged to be better suited for GAS proposals that would be complementary to the CCS rather than included directly in the CCS design.  
The ROTAC encourages the community to continue to pursue GAS programs that benefit from the CCS data, but may require additional observations. 

Conserving the valuable resource of the GA time was a high priority of the ROTAC, which considered the impact of the CCS allocations on the amount of time available for these programs.
When assessing the CCS designs, the ROTAC took into account which potential science cases outside of the survey requirements could be addressed with a separate GAS program optimized by the community. In general, the underguide options for the CCS options were felt to be sufficient to meet the Roman science requirements for these individual surveys, but would not produce datasets with the most expansive scientific value for the community, while the nominal designs would produce data applicable to several of the additional science cases provided to the committees via white papers (e.g., Appendix~A of the GBTDS report in \autoref{app:gbtds}).  Naturally, the overguide options were able to address not only more of the core science but also additional ancillary science with the additional time allocation.

As described in \autoref{sec:ccs}, the ROTAC felt that the nominal designs for the HLWAS and HLTDS nicely balanced the amount of time above the underguide with the benefits of the additional data to the community; these additional data could be achieved most easily through providing those surveys with the nominal time allocation, since any separate program would be a significantly less efficient way to achieve that critical science. However, the GBTDS overguide option adds significantly more community value to the survey, as determined through the submitted white papers, even at slight expense to the total planet yield (\autoref{sec:gbtds}). Thus, the ROTAC is recommending the GBTDS overguide allocation. This recommendation still enables the mission to meet its goal of GAS time allocation while providing a large amount of additional time-domain data, with a wealth of GA science capabilities, to the community.  

In addition to total time allocation, the scheduling considerations of the CCS and their impact on the GAS were also taken into account. The ROTAC considered the value of having GAS observations early in the mission (e.g., in the first year), which is also in heavy demand by the CCS. The ROTAC recommends preserving some room for GAS programs that require long time baselines (e.g., proper motion studies) in the schedule; however, it is clear and justified that the CCS all need to obtain early observations to verify that their long-term strategies, which stretch across the mission, will be viable. The ROTAC was provided with some preliminary mission schedules (e.g., \autoref{fig:observing_schedule}) to confirm that GAS time will be available interspersed across the full 5-year mission given our recommendations for the CCS.

\section{Process \& Timeline}

\subsection{Committee Empanelment}
In July 2024, the Roman Space Telescope team sent out a call\footnote{\url{https://roman.gsfc.nasa.gov/science/roman_time_allocation_committee.html}} for self-nominations of individuals willing to serve on the Roman Observations Time Allocation Committee (ROTAC). Over fifty nominations from around the world were received, and in December 2024 the committee membership was announced.  The committee is composed of thirteen scientists: ten community representatives with expertise covering a broad range of astrophysics, and one co-chair from each of the three Core Community Survey Definition Committees.  The ROTAC is co-chaired by Dr. Gail Zasowski (The University of Utah) and Dr. Saurabh W.~Jha (Rutgers University). The rest of the committee includes: \vspace{-6pt}
\begin{itemize}[nolistsep]
    \item Laura Chomiuk, Michigan State University
    \item Xiaohui Fan, University of Arizona
    \item Ryan Hickox, Dartmouth College
    \item Dan Huber, University of Hawai`i 
    \item Eamonn Kerins, University of Manchester
    \item Chip Kobulnicky, University of Wyoming
    \item Tod Lauer, NOIRLab
    \item Masao Sako, University of Pennsylvania
    \item Alice Shapley, University of California, Los Angeles
    \item Denise Stephens, Brigham Young University
    \item David Weinberg, The Ohio State University
    \item Ben Williams, University of Washington
\end{itemize}

The ROTAC was charged with reviewing the implementation reports of the three core community-defined surveys and advising the Roman Project on the implementation of each survey. The ROTAC was also charged with considering the balance between these CCS and the future PI-led suite of General Astrophysics Surveys (GAS). To accomplish these tasks, the ROTAC was instructed to: \vspace{-6pt}
\begin{itemize}[nolistsep]
    \item Become familiar with the Roman science requirements, including the Roman Mission objectives in cosmology and exoplanet demographics that were to be met with the CCS portfolio.
    \item Review the input from the CCS DCs and discuss the recommended implementation options, with the DC representatives as needed to fully understand the recommendations.
    \item Represent the interests of the scientists who will make use of Roman through both the CCS and the GAS, when considering the legacy value of the recommended CCS implementation options and potential impacts of the CCS time allocation on the PI-led GAS.
    \item Work with the Roman Project at NASA/GSFC and the Roman Science Centers to understand any scheduling conflicts or other operational impacts among the recommended CCS designs. If necessary, the ROTAC was to provide advice to the Roman Project at NASA/GSFC on adjudicating any conflicts among the CCS.
    \item Present the ROTAC recommendations to the Roman Project in a written form suitable for release to the scientific community.
\end{itemize}

\subsection{Meeting Cadence and Format}
The ROTAC met twice in December~2024 for orientation and planning purposes, and then convened each week from late January through early April~2025. The committee met through Zoom, and the day and time of the meeting was alternated every other week to allow all members of the ROTAC to be able to attend synchronously at least once every two weeks. Meeting minutes and recordings were made available to the full committee to review.

The ROTAC chairs guided discussions of how science return would scale in the underguide, nominal, and overguide time allocation scenarios, and of practical considerations to increase the total science return and efficiency of the unified WFI (CCS+GAS) program.  Ongoing real-time input from representatives of the Core Community Surveys greatly facilitated these discussions. 
The ROTAC progress was guided in part by a series of asynchronous polls on the science cases and implementation options that highlighted where the committee shared general agreement and where focused discussion would be useful. In general, these ROTAC poll responses showed a broad consensus that shaped the ultimate recommendations for the allocations for each survey.

\subsection{Hybrid Meeting and Finalization of Recommendations}

A hybrid ROTAC meeting was held March 20--21, 2025, at the Space Telescope Science Institute in Baltimore, MD, with roughly 75\% attending in person and the remainder online. 
The session started with a high level overview of the Roman Project and a reminder of the ROTAC's charge to recommend an optimal balance between the CCS and the GAS, which together are allocated 1527~days of Roman observing time. Over the course of the two days, the ROTAC  received presentations and updates from various technical teams, in particular from the SOC on progress to test the scheduling of various CCS combinations into the APT software, which had been requested earlier by the ROTAC. The SOC also shared a visualizations of the multiyear scheduling plan (e.g., \autoref{fig:observing_schedule}), to highlight relevant conflicts and scheduling issues. This plan did not reveal any significant pinch points that required resolution by the ROTAC.

After a significant amount of time spent discussing scheduling considerations for the GBTDS, the ROTAC requested and received a status report from the Galactic Plane Survey definition committee. At that point, the committee felt it had reached consensus on its recommendations.

\section{Considerations for the Recommendations}

The ROTAC's recommendations are based on a wide array of information supplied by the reports of the DCs and by the the Roman Project Office and Survey Operations team. These include the array of science objectives across the core surveys, trade-offs in those objectives for differing survey designs, and the potential for scheduling conflicts between the surveys. Roman's capabilities offer ample opportunity for ground-breaking surveys across many areas of astrophysics. The ROTAC recommendations favor CCS designs that offer maximal potential for additional astrophysics. The ROTAC has also been keen to safeguard observing windows that will encourage ambitious GAS proposals. Given all of these factors, inevitably there remain some items that cannot be optimized at this stage; we describe below those that arose during our discussions with the highest potential impact.

\subsection{Ongoing Survey Optimizations}

The ROTAC recommendations are designed with three objectives. Firstly, they include provision for specific elements within each of the CCS that we regard as scientifically excellent and that exploit Roman's powerful survey capability. Secondly, they set out recommended overall time allocations for each of the CCS. Thirdly, they establish the overall time envelope for CCS activity and therefore the time available for GAS programs.

The ROTAC recommendations recognize the high scientific value of different elements comprising each of the CCS. These include the pilot, extended and core components specified in the nominal HLTDS design, the medium, wide and deep tiers of the nominal HLWAS, and the high cadence Galactic Center and microlensing fields of the overguide GBTDS. 
The ROTAC recommends that all of these CCS elements should be present within the final survey designs. However, within the bounds of the overall recommended time allocations, we leave the final choices of filter strategy and precise field placement as matters for the Roman mission in consultation with the respective experts (e.g., the Project Infrastructure Teams and DCs). 

The Science Operations Center presented to the ROTAC some early modeling of the Roman observation schedule that included nominal HLWAS, nominal HLTDS, overguide GBTDS, and GPS components (e.g., \autoref{fig:observing_schedule}). Provisional analysis indicates that these options can all be accommodated with significant flexibility for a range of GAS program designs. The ROTAC is satisfied that the provisional scheduling of the CCS and the GPS leave sufficient observing windows of varying duration to facilitate high-impact Roman GAS programs. We acknowledge and stress, however, that modeling the Roman observing schedule remains an ongoing and iterative process.

\subsection{External Data Requirements}

All three core surveys will benefit from additional data from other projects. The GBTDS does not have any formal requirement for ancillary data but will benefit from early observations made by Hubble and Euclid. Additional simultaneous observations from Rubin, PRIME, and Subaru, as well as from the OGLE, MOA, and KMTNet microlensing surveys, will also help extend and improve the GBTDS science potential.

To secure its scientific objectives, the HLWAS will require optical photometry from Rubin, as well as ground-based calibration spectroscopy within the deep fields. The overall footprint design also includes overlap with other large-area surveys.

The HLTDS will prioritize Roman's unique capabilities to precisely measure higher redshift SNe, but the full cosmological measurement will require a high-quality sample of Type~Ia SNe across the full redshift range from $z = 0$ to
$z \approx 2$. For the intermediate redshifts ($z \sim 0.5$), SNe~Ia from the Rubin LSST Deep Drilling Fields will be crucial to satisfy the HLTDS cosmology requirements, while surveys targeting SNe~Ia in the low-redshift Hubble flow will also play a key role.

In its considerations, the ROTAC assumed that all ancillary data necessary to fulfill Roman CCS science requirements will be available to the entire science community, just as Roman data itself will be. The ROTAC did not take into account the availability of data that may improve Roman science, but which is not critical to satisfy core requirements.

\subsection{Retaining Flexibility for General Astrophysics}

While the core science objectives for Roman are well defined, and the designs for the CCS are relatively mature, we do not yet know about GAS programs that Roman will undertake, with the exception of the GPS. This uncertainty argues for maintaining flexibility in CCS scheduling. One example of this is the low-cadence seasons of the GBTDS. The original overguide design recommended by the GBTDS DC proposed a three-day observing cadence during those bulge seasons not scheduled for the high-cadence GBTDS observations. This low-cadence scheduling will be important for long-timescale light-curve characterization, but it may also make it very difficult to accommodate other GAS programs, especially any focusing on the bulge region. It may also cause significant slew overheads resulting from clashing GBTDS and HLTDS cadences, indirectly impacting the GAS. These negative outcomes should be avoided, for example by adjusting the observing strategy of the low-cadence GBTDS (\autoref{sec:gbtds} above). 

As also described above, we recommend that two days during the middle of the first GBTDS high-cadence bulge season be used in GPS observations (part of the GAS) for the wider Galactic bulge region. These initial observations will establish an essential early baseline for subsequent GPS passes to enable the highest-quality Galactic bulge astrometric studies.

\subsection{Recommendations amid Change and Uncertainty}
\label{sec:uncertainty}

The ROTAC emphasizes that our recommendations made at this point in time are --- as they must be --- based on a current snapshot of an evolving scientific landscape. Mission scheduling and survey optimization remain active and ongoing activities that may require further tuning of the allocations put forward in our report.

Additionally, the CCS represent ambitious, multi-year commitments. We expect that during the remaining preparatory and survey execution phases, new science results may emerge from Roman, or from other surveys, that argue for adaptation in CCS strategy for Roman to capitalize best on new knowledge. The ROTAC also envisages that mid-mission reviews of the performance of the surveys may also prompt revisions of survey priorities and strategy and notes there is an expectation that GAS observations would play a larger role in a potential Roman extended mission. The ROTAC is keen to ensure that its recommendations leave scope for survey modifications that are responsive to an evolving mission and scientific landscape.

An example comes from the design of the GBTDS survey. The original design assumed a fiducial planet mass function with an increasing number of low mass planets per host until saturation at 2~planets per~dex$^2$ in logarithmic mass and host separation (Section~3.2.2 of the GBTDS DC report; \autoref{app:gbtds}). 
However, updated microlensing measurements indicate that the saturation level for the lowest mass planets may be at, or below, 0.5~planets per star per~dex$^2$. The GBTDS science requirement for the number of low mass planet detections to be made by Roman was motivated by the assumed fiducial mass function; it is not reachable in practice if the updated mass function limit accurately describes the Universe. Guided by the Roman mission office, the ROTAC interpreted the GBTDS science requirement as representing the design goal for Roman if the earlier fiducial mass function represents reality. We are satisfied that our recommended design option comfortably satisfies that goal. 

The fact that Roman CCS design requirements deployed metrics within regions of ill-defined discovery space that have subsequently needed revision was not seen as a flaw or a concern by the ROTAC. Rather, it serves to highlight that Roman is a mission that is designed to explore regions of discovery space that are highly uncertain. {\bf It underscores just why we need Roman, and why Roman will be a revolutionary facility for exoplanetary science, cosmology, and astrophysics at all the scales in between.}

\clearpage

\appendix

\section{Global Contributions to the Roman WFI Observing Plan}
\label{app:world_contributions}
These maps highlight the institutions of scientists who contributed to Roman white papers or science pitches, and those who served on the Project Infrastructure Teams (PITs), CCS and GPS DCs, and the ROTAC. The wide distribution across six continents demonstrates the global enthusiasm for this mission.
\begin{figure}[!h]
\includegraphics[width=\textwidth, trim=0cm 3.9cm 0cm 3.9cm, clip]{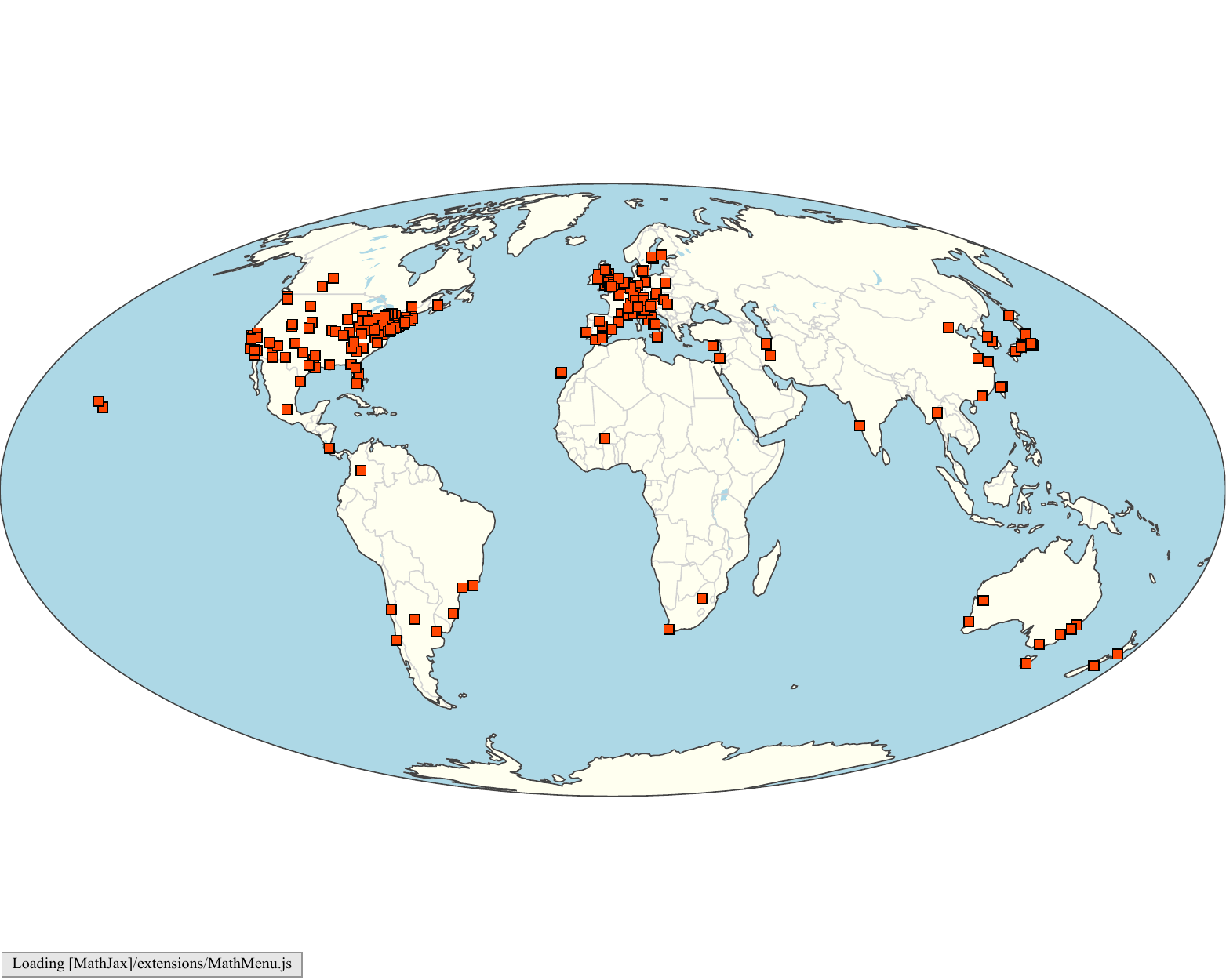}
\includegraphics[width=0.45\textwidth]{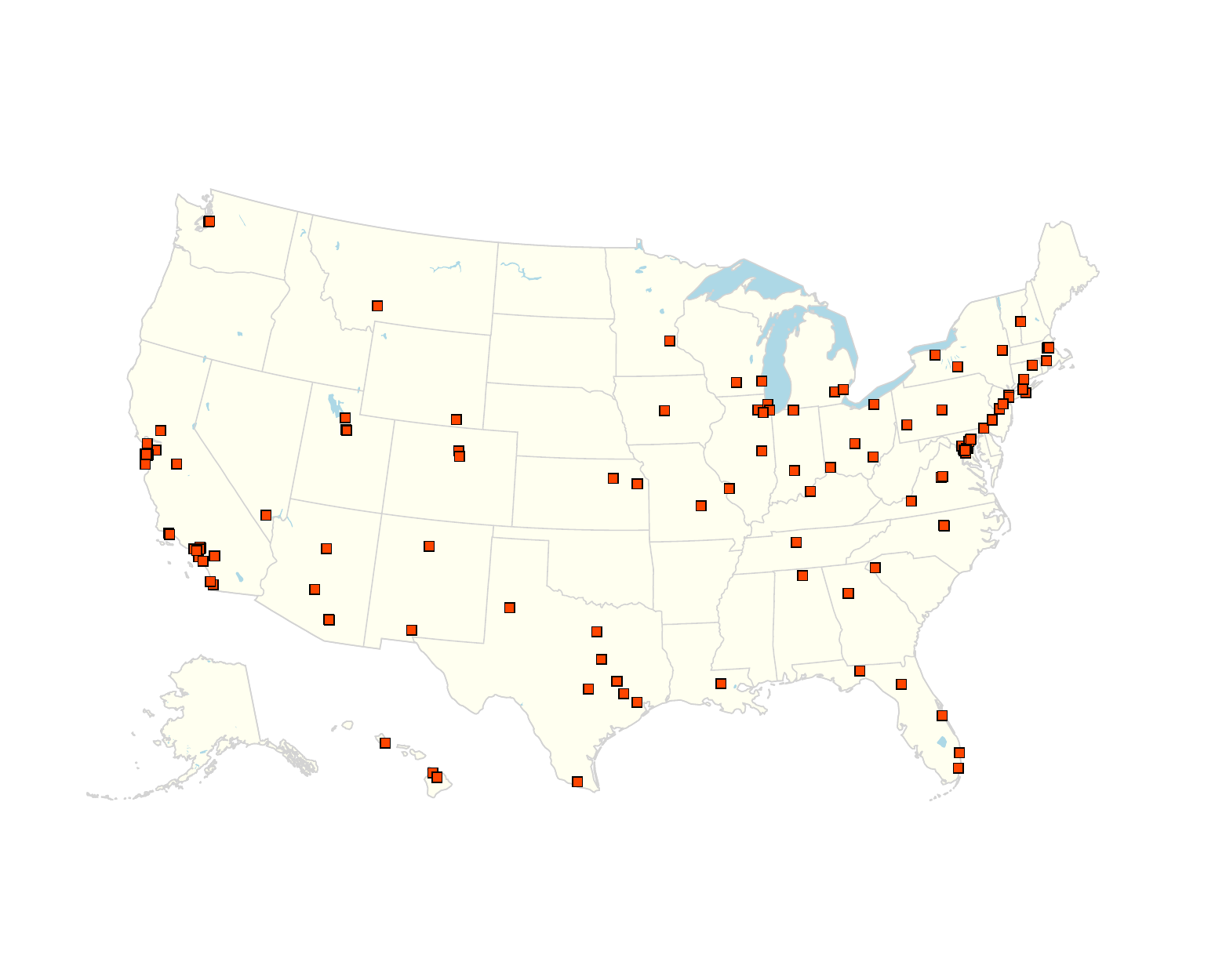}
\includegraphics[width=0.45\textwidth]{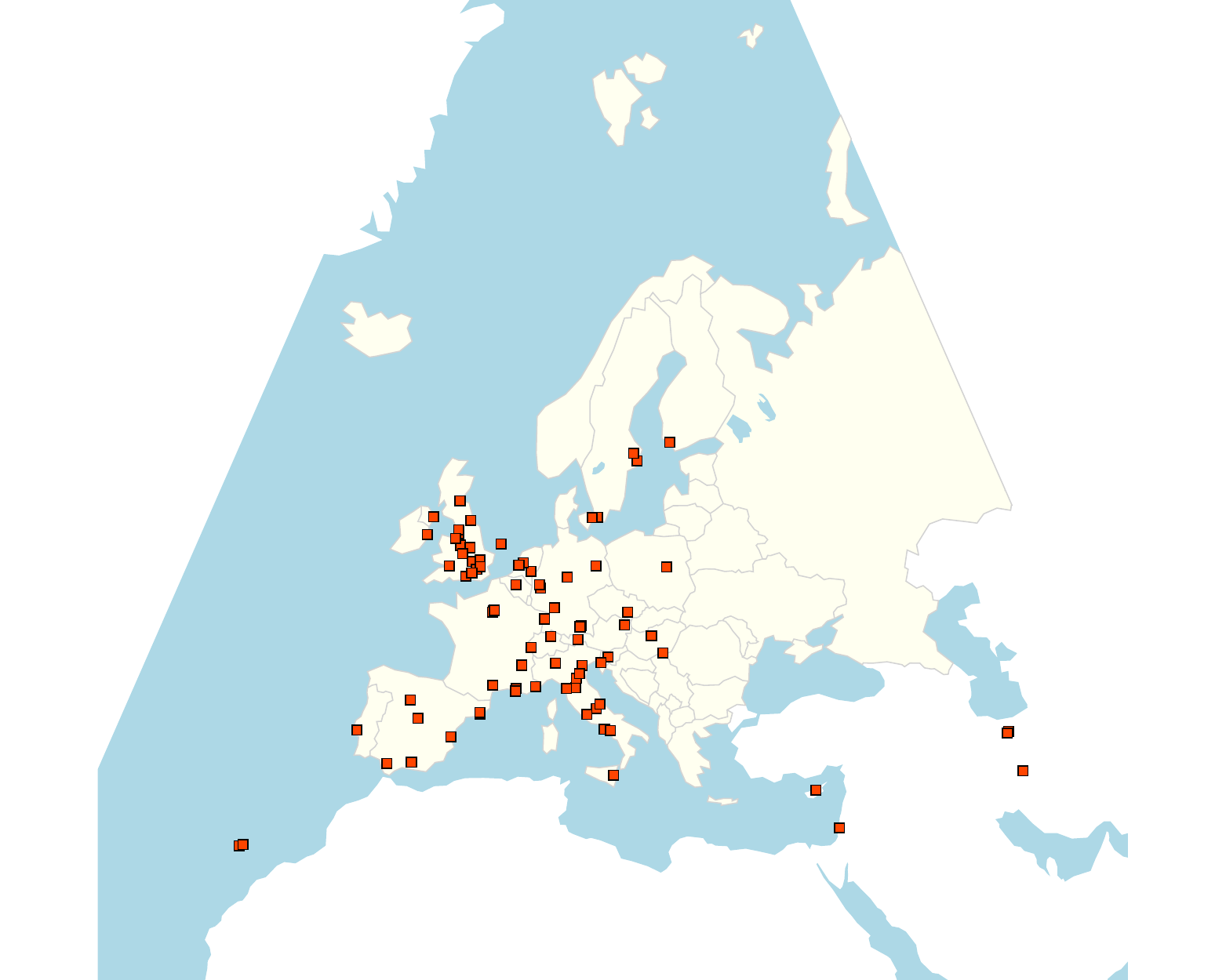}
\caption{Institutions of individuals who contributed to white papers and science pitches, and who served on the PITs, the DCs, and the ROTAC. Figure credit: Gail~Zasowski; additional data provided by Leslie~Beauchamp, Jenna~Ballard, and Julie~McEnery.}
\label{fig:insts}
\end{figure}
\clearpage

\section{Works Cited in This Report and the CCS DC Reports}
\label{app:refs}
We include here all references from the DC reports in \autoref{app:dc_reports}.
\nocite{*}

\begingroup
\renewcommand{\section}[2]{}%
\bibliographystyle{yahapj}
\bibliography{hlwas,HLTDS,gbtds,main}{}
\endgroup
\clearpage

\section{Core Community Survey Definition Committee Reports}
\label{app:dc_reports}

These reports from the CCS DCs were released in December~2024.

\subsection{High Latitude Wide Area Survey}
\label{app:hlwas}
\includepdf[pages=-]{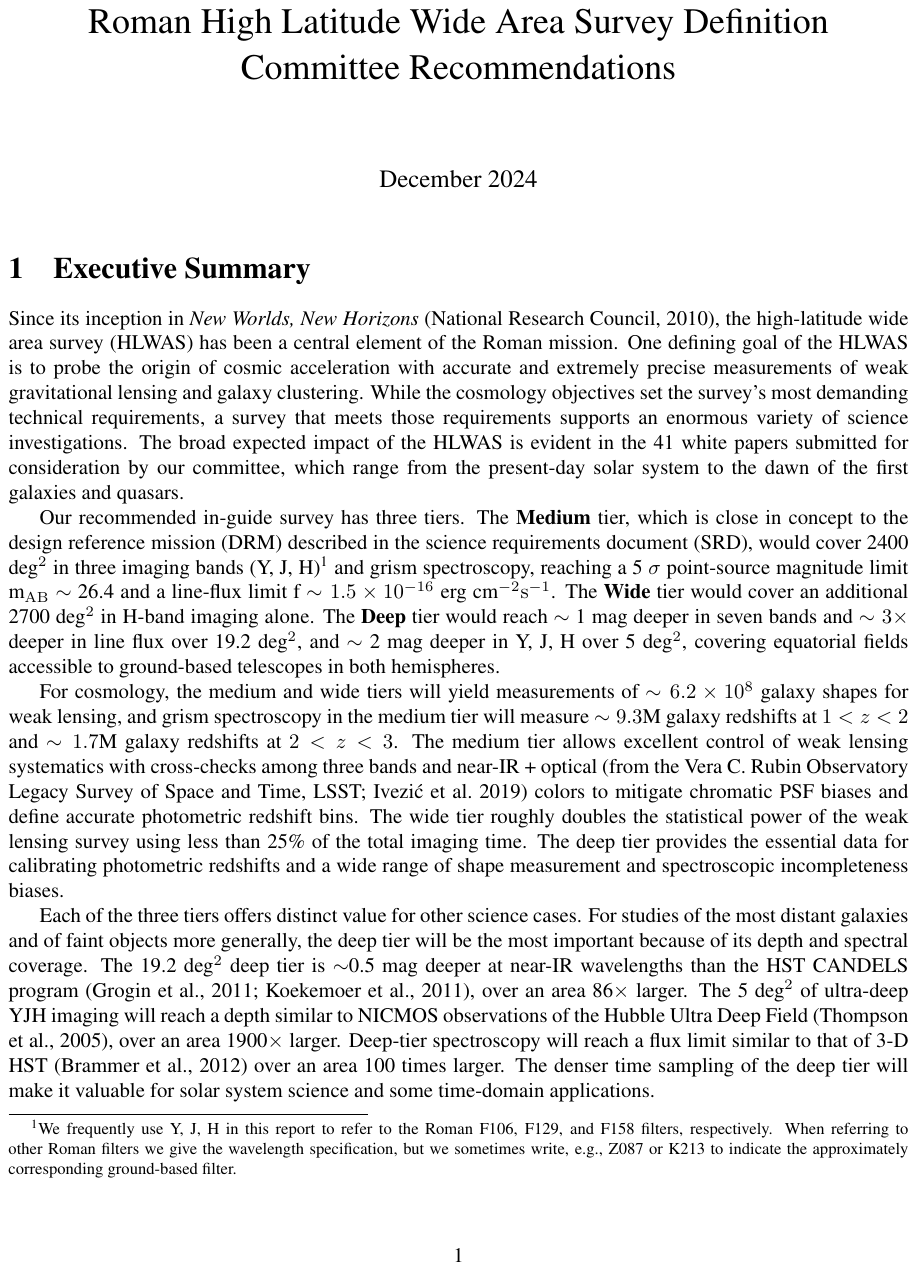}
\subsection{High Latitude Time Domain Survey}
\label{app:hltds}
\includepdf[pages=-]{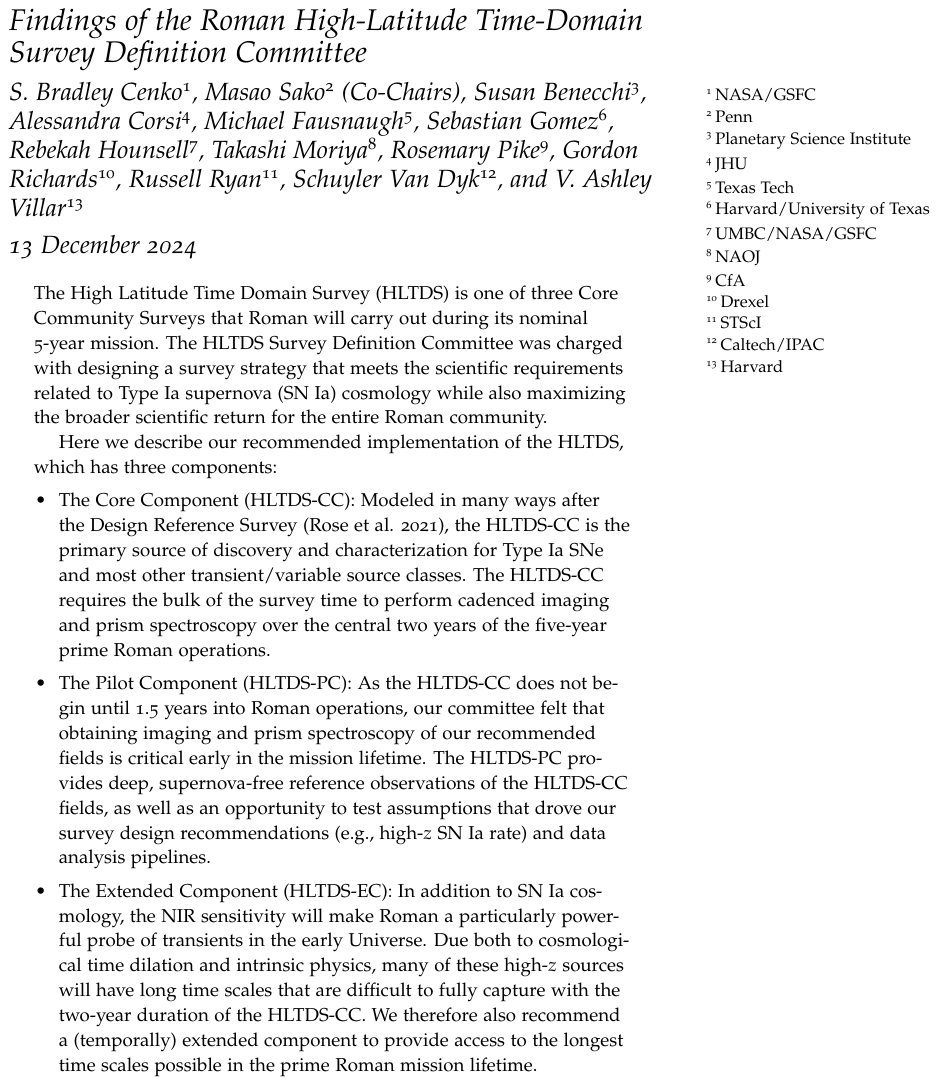}
\subsection{Galactic Bulge Time Domain Survey}
\label{app:gbtds}
\includepdf[pages=-]{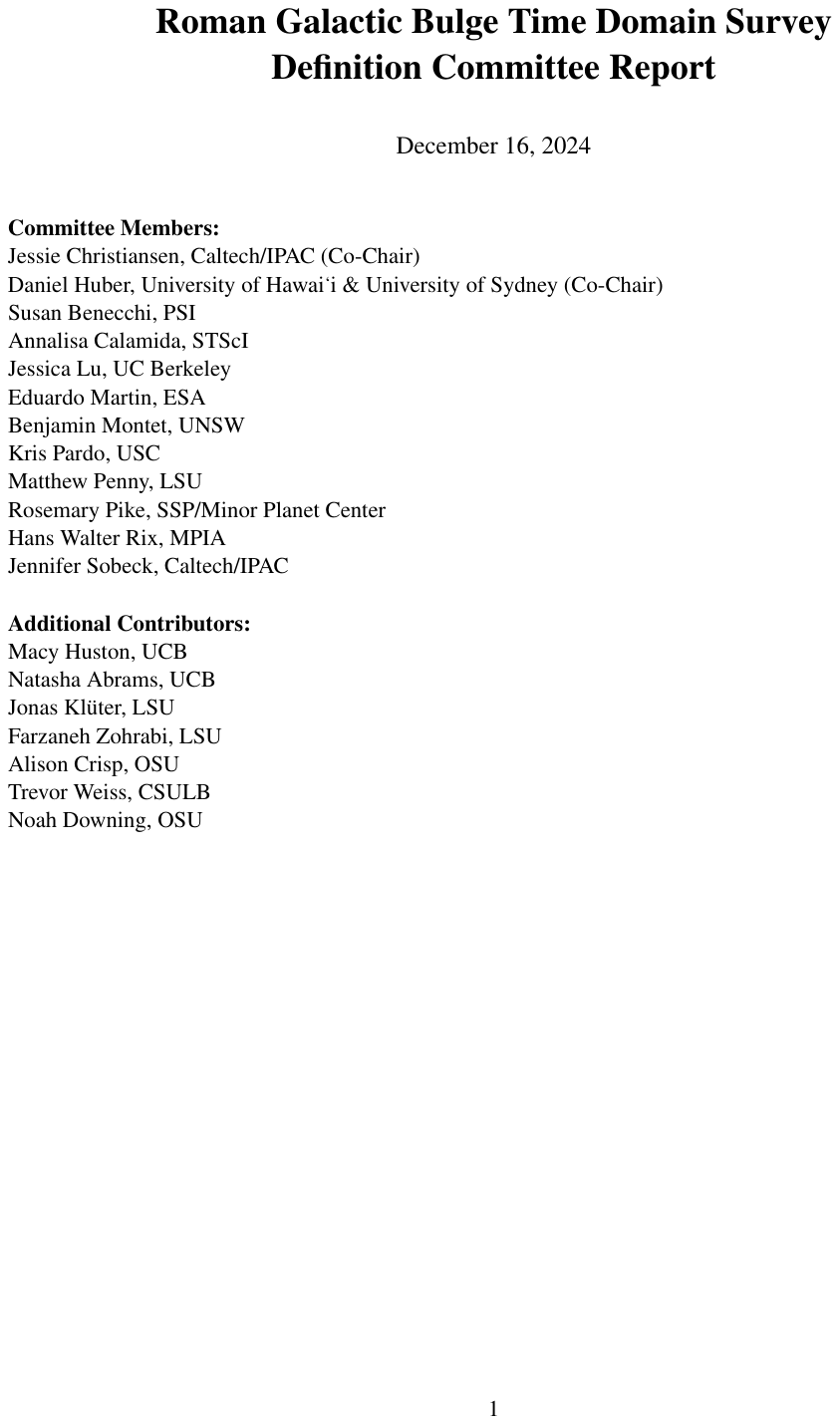}

\end{document}